\documentclass[trackchanges,twocolumn,twocolappendix]{aastex701}
\usepackage{placeins}

\begin{document}

\title{Variable X-ray flux may explain variable helium escape from the habitable-zone exoplanet LHS~1140b}

\author[orcid=0000-0002-8466-5469]{Collin Cherubim}
\altaffiliation{NHFP Sagan Fellow}
\affiliation{Department of Astronomy \& Astrophysics, University of Chicago, Chicago, IL 60637, USA}
\email[show]{collinc@uchicago.edu}  

\author[orcid=0000-0001-7296-3533]{Tim Cunningham} 
\altaffiliation{NHFP Hubble Fellow}
\affiliation{Center for Astrophysics, Harvard \& Smithsonian, 60 Garden St., Cambridge, MA 02138, USA.}
\email{timothy.cunningham@cfa.harvard.edu}

\author[orcid=0000-0001-7730-2240]{Jason A. Dittmann} 
\affiliation{Department of Astronomy, University of Florida, 1772 Stadium Rd, Gainesville, FL 32611, USA.}
\email{jasondittmann@ufl.edu}







\begin{abstract}

A central goal in exoplanet science is to understand the conditions under which potentially habitable worlds retain atmospheres. This question is of special interest for worlds orbiting M stars, whose habitable-zone planets are more readily observable, but which also emit prolonged high-energy radiation that can erode planetary atmospheres. For one such world, LHS\,1140b, a rocky planet orbiting in the habitable zone of an inactive star, time-variable helium absorption has been observed as the planet transited in front of its host star, consistent with atmospheric escape. We present X-ray/ultraviolet flux observations with the \textit{XMM-Newton} space telescope, which serve to test whether the observed helium absorption variability can be explained by stellar XUV flux variability. We detect X-ray flux variability with an amplitude of $F_\mathrm{max}/F_\mathrm{min} = 8.5^{+9.4}_{-4.4}$ within the 18-hour observing window. We do not detect variability in the UV or visible flux. The integrated X-ray flux is fully consistent with the X-ray flux measured in 2018, suggesting stable XUV flux over years-long timescales. We find that the 2018 observation was not sensitive to the level of variability observed in 2026. We explore several models that fit both datasets and conclude that short-term XUV variability and/or small flares are more likely to explain the helium variability than long-term XUV variability. Simultaneous observations of the planet-transit in the near-infrared and stellar activity indicators like XUV flux are critical to confirm the link between the observed helium signal variability and the observed X-ray variability.

\end{abstract}

\section{Introduction}

X-ray and extreme ultraviolet (XUV) stellar irradiation heats exoplanet atmospheres orbiting close to their stars, driving hydrodynamic atmospheric escape. This mechanism dominates planetary atmospheric evolution and the sculpting of demographic features like the exoplanet radius valley and hot Neptune desert \citep{szabo_2011, Fulton_2017, Owen_2019}. XUV-driven escape is directly observed for transiting exoplanets via extended neutral hydrogen (H) winds via the Ly-alpha line and the metastable helium (He) triplet \citep{Oklopcic_2018, Owen_2023b}. While these detections have been mostly limited to giant planets and sub-Neptunes, the recent detection of escaping He from the super-Earth LHS~1140b presents an opportunity to study the link between XUV flux and atmospheric escape for an exoplanet that is more Earth-like in its bulk composition and temperature \citep{Cherubim_2026}.

LHS\,1140b orbits an old ($> 3$ Gyr), quiescent M4.5-class star located 14.96 ± 0.01 parsecs away ~\citep{Dittmann_2017, Medina_2022, Pass_2024, Cadieux_2024a}. With an orbital period of 24.7 days, it receives 42\% of the stellar irradiation received by Earth, giving it an equilibrium temperature of $T_\mathrm{eq}$ = 226 ± 4 K, which places it in the liquid-water habitable zone ~\citep{Dittmann_2017, Cadieux_2024a}.

\cite{Cherubim_2026} used the high-resolution WINERED spectrograph ($R$ = 68,000) mounted on the Magellan/Clay telescope at Las Campanas Observatory to observe LHS\,1140b in two epochs: Sep. 23 2024 and Sep. 28 2025. He absorption was observed when the planet passed in front of the host star (i.e., transited) in 2024. The mass-loss rate of He was inferred assuming a semi-empirical stellar XUV flux based on an X-ray measurement taken by \textit{XMM-Newton} on December 21, 2018 (PI: Dittmann; Obs ID: 0822600101). 

On Sep. 28 2025, \cite{Cherubim_2026} observed one additional transit of LHS~1140b with the same instrument setup and detected no He absorption and an upper limit of $\sim$0.6\%, about half that measured in 2024. 
A similar upper limit was placed on planetary He absorption for each of four planetary transits observed with the NIRISS instrument on the \textit{James Webb Space Telescope} (\textit{JWST}) on Dec. 1 2023, Dec. 25 2023, Jul. 26 2025, and Jul. 23 2026 \citep{Radica_2026, Bennett_2026, Gressier_2026}. These observations show no evidence of He absorption, though none were contemporaneous with the detected signal on Sep. 23 2024, with the closest in time being nine months prior.

Stellar XUV irradiation is known to vary in time \citep[e.g.,][]{Hathaway_2015, Pillitteri_2026}, and planetary escape signatures, including metastable He absorption, are known to vary in time \citep{Etangs_2012, Guilluy_2020, Zhang_2022, Masson_2024, Levine_2024, Taylor_2025, Allart_2025, Saidel_2026}.
As stellar XUV irradiation is known to drive atmospheric escape, we hypothesize that the variability in the LHS~1140 He absorption lines observed during the transit of LHS~1140b is due to variable stellar XUV flux that drives variable atmospheric escape from the planet. Here we present results from X-ray and UV observations with \textit{XMM-Newton} to test this hypothesis over both yearly (2018 -- 2026) and hourly (18 hr) timescales.

\section{Observation and data analysis}
\label{sec:analysis}

LHS~1140 was targeted with \textit{XMM-Newton} \citep{jansen2001} on January 12, 2026 (PI: Cherubim, Cunningham; Obs ID: 0982690101). The observation was carried out in a single visit for a total duration of 64\,ks. The observation was interrupted twice and divided into three separate exposures with durations of 1.22, 5.60, and 54.43 ks. The two shorter exposures are dominated by high flaring particle background and are excluded from the data analysis. The source was detected in the three detectors of the European Photon Imaging Camera (EPIC): PN \citep{struder2001}, MOS\,1 and MOS\,2 \citep{turner2001}. We reduced the data using the Science Analysis Software (SAS; v22.0.0), designed for the reduction of \textit{XMM-Newton} data \citep{gabriel2004-sas}. We also collected seven 4.4-ks exposures with the Optical Monitor (OM) in Fast Mode with both the UVW1 and V filters.

We filtered the event lists, which catalog the energy and arrival time of each detected photon/particle, using the \texttt{evselect} routine with the default pixel pattern selection for the PN and MOS detectors, including photon energies between 0.2--10.0\,keV. We defined Good Time Intervals (GTI) using the \texttt{tabgtigen} routine with threshold rates of 0.4, 0.16, and 0.2 counts\,s$^{-1}$ for PN, MOS\,1, and MOS\,2, respectively, determined from visual inspection of the total count rate over time. The resulting total effective exposure times for PN, MOS\,1, and MOS\,2 were 44.9, 49.5, and 49.7 ks, respectively. We then used the \texttt{epiclccorr} routine to correct for events lost through inefficiencies in the mirror-detector system such as vignetting, bad pixels, chip gaps, point spread function (PSF), and quantum efficiency. Finally, we extracted spectra for the source and a nearby background region using \texttt{evselect}.

The source suffers from contamination by a brighter nearby source (Gaia DR3 2371032989200665984, Gmag\,$=$\,16.7). In the 2018 epoch, LHS~1140 had a separation of $\sim$14 arcsec from the nearby contaminant. Our follow-up observation benefited from better separation from the contaminant, as LHS~1140 has relatively high proper motion (0.67 arcsec\,yr$^{-1}$), while the contaminant does not. From the effect of proper motion, in our 2026 observation, LHS~1140 was $\sim$5 arcsec farther from the contaminant. We performed point spread function (PSF) photometry on the event lists to estimate the source aperture sizes for LHS~1140 and the contaminant in order to minimize contamination. This procedure revealed an optimal source aperture radius of 10 arcsec for both the source and contaminant (Figure \ref{fig:apertures}. We estimate that the contaminant is likely to contribute $\sim$5\% of the flux in the 10-arcsec source aperture of LHS~1140 based on the overlap of the two PSFs.

To measure the X-ray flux from LHS~1140, we used the Bayesian X-ray Analysis (BXA) package \citep{buchner2016-bxa}. The BXA package combines standard X-ray spectral models with the nested sampling algorithm \texttt{UltraNest} \citep{buchner2019-UltraNest}. BXA generates posterior distributions for model parameter estimation and calculates marginal likelihoods for model comparison. We used the X-ray Spectral Fitting Package (\texttt{XSPEC}) to model the X-ray spectra in our analysis with BXA \citep{arnaud1996XSPEC}. We included two components in our \texttt{XSPEC} model: The Tübingen-Boulder ISM absorption model, which models X-ray absorption by the interstellar medium (ISM); and the Astrophysical Plasma Emission Code (APEC), which is an isothermal, optically-thin plasma model. At the distance to LHS~1140, 14.96\,$\pm$\,0.01 pc, the ISM hydrogen column density is expected to be approximately $N_\mathrm{H} = 1.45 \times 10^{18}$ cm$^{-2}$, which is expected to have a negligible effect in the EPIC passbands \citep{redfield2000}. 

In our spectral analysis with BXA, we simultaneously fit all EPIC data across two spectra: one PN spectrum and one spectrum for both MOS datasets, which were merged using the SAS routine \texttt{epicspeccombine}. We fit for four parameters between a two-component, two-temperature APEC model, combined additively, for each of the two spectra: two APEC temperatures ($kT_\mathrm{1}$, $kT_\mathrm{2}$) and two normalization parameters ($n_\mathrm{1}$ and $n_\mathrm{2}$). The normalization parameters were fit in log space. Uniform priors were adopted for the APEC temperatures, $kT_\mathrm{1} = 0.05-0.5$ keV and $kT_\mathrm{2} = 0.5-3.0$ keV, and for both normalization parameters $n_\mathrm{1}, n_\mathrm{2} = 1 \times 10^{-7} - 1 \times 10^{-4}$. The spectra were fit using C-statistics due to low source counts \citep[79 in PN, 14 in MOS\,1 and 17 in MOS\,2][]{cash1979}. We adopted solar elemental abundances in our model from \cite{aslpund2009}. The abundance parameter was fixed at unity and the redshift parameter was fixed at zero and neither was sampled in the model fit. The ISM hydrogen column density was fixed at $N_\mathrm{H} = 1.45 \times 10^{18}$ cm$^{-2}$, and our results were insensitive to this choice.

\begin{figure*}[t!]
\centering
\includegraphics[width=0.8\textwidth]{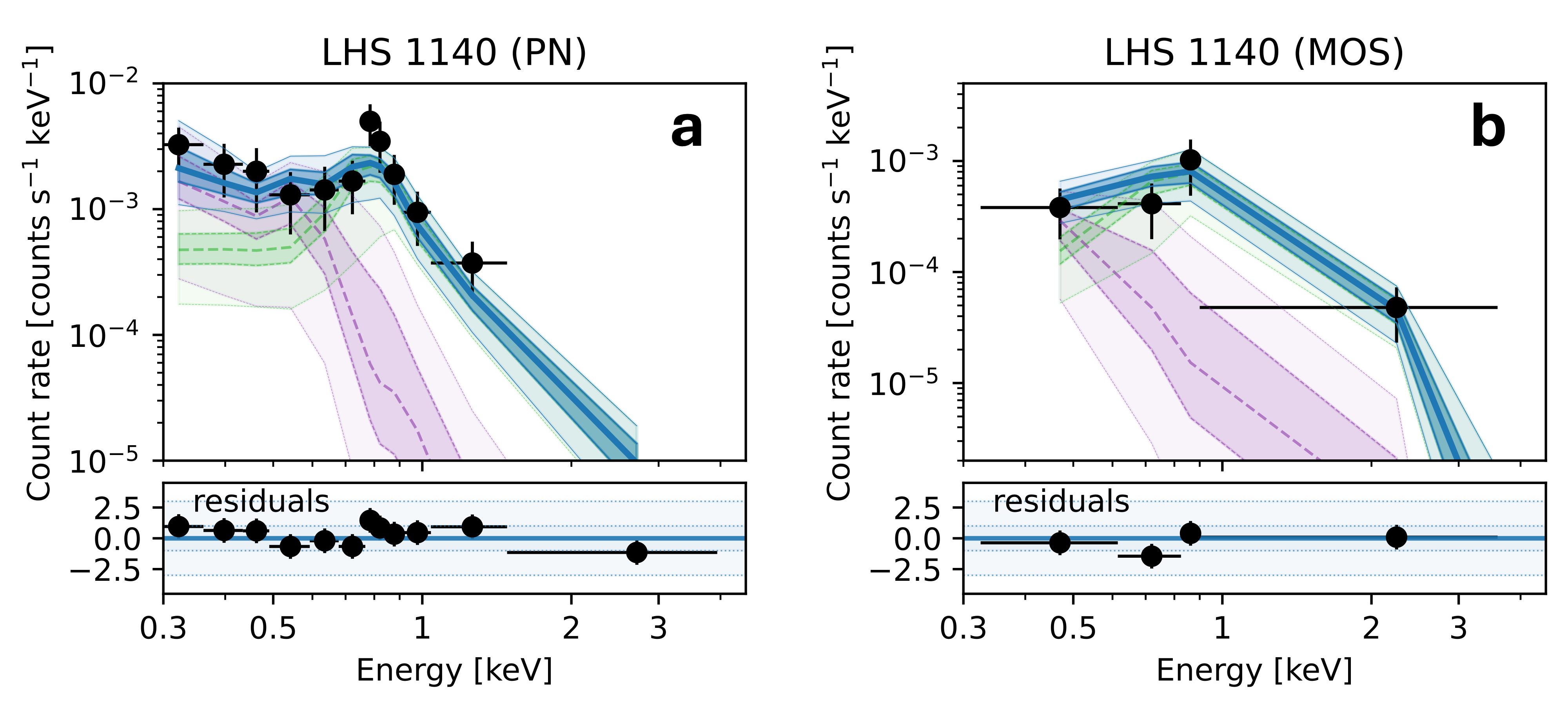}
\caption{X-ray spectral analysis. The best-fit models, and residuals are plotted for the EPIC-PN detector (a) and for the combined EPIC-MOS detectors (b). The soft APEC component is plotted in purple and the hard component is plotted in green. The combined model is shown in blue. Dashed lines indicate median values while shaded areas indicate standard deviation.}
\label{fig:spectra}
\end{figure*}

\begin{table}[t]
\def\arraystretch{1.2}
\centering
\caption{X-ray flux, $F_{\rm X}\ [10^{-15}\,\mathrm{erg\,cm^{-2}\,s^{-1}}]$, and best-fit temperatures, $kT$, for LHS~1140 from a four-parameter fit of a two-temperature optically-thin plasma APEC model with BXA ($1\sigma$ errors; 90\% confidence limit in parentheses). The posteriors for all four parameters are shown in Figure \ref{fig:corner}. The 2018 values were determined using the same data pipeline reduction by \cite{Cherubim_2026}.}
\begin{tabular}{ll}
\hline
\textbf{This work} \\
$F_{\rm{X}} (0.25-2.0\ \mathrm{keV})$ 
    & $3.15^{+0.53}_{-0.65}$ (2.31, 4.26) \\
$F_{\rm{X}} (0.25-10.0\ \mathrm{keV})$ 
    & $3.18^{+0.54}_{-0.66}$ (2.32, 4.29) \\
$kT_1$ [keV] 
    & $0.11^{+0.05}_{-0.03}$ (0.07, 0.19) \\
$kT_2$ [keV] 
    & $0.54^{+0.12}_{-0.15}$ (0.33, 0.75) \\
\hline
\textbf{2018\footnote{\cite{Cherubim_2026}}} \\
$F_{\rm{X}} (0.25-2.0\ \mathrm{keV})$ 
    & $3.07^{+0.45}_{-0.41}$ (2.40, 3.83) \\
$F_{\rm{X}} (0.25-10.0\ \mathrm{keV})$ 
    & $3.15^{+0.50}_{-0.42}$ (2.44, 3.99) \\
$kT_1$ [keV] 
    & $0.22^{+0.019}_{-0.017}$ (0.18, 0.25) \\
$kT_2$ [keV] 
    & $3.7^{+1.6}_{-1.8}$ (1.6, 5.6) \\
\label{table1}
\end{tabular}
\end{table}

\section{Results}

\begin{figure*}[t]
\centering
\includegraphics[width=\textwidth]{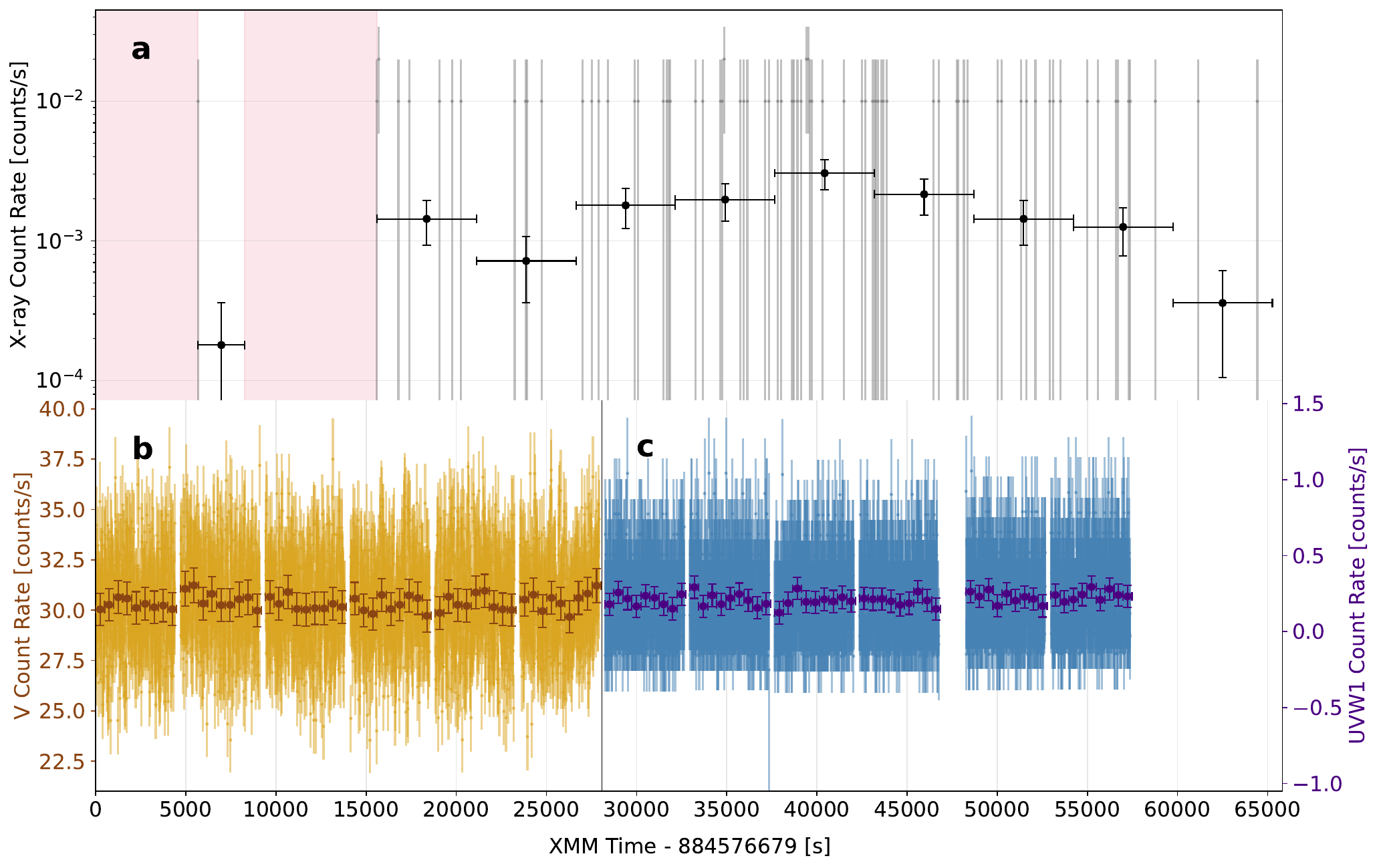}
\caption{Light curves for EPIC PN (a), OM V filter (b), and OM UVW1 filter (c). Shaded pink regions show omitted periods of high background noise and/or interrupted data collection. These regions are not included in the spectral modeling. In panel a, gray points show data binned to 100 s with error bars generated from the XMM-Newton SAS pipeline. The absence of lines indicates zero counts. Black points show 5600 s bins, which include a minimum of one count for the source aperture and errors added in quadrature. In panels b and c, smaller points have 10 s bins with error bars generated from the PPS pipeline and larger points have 500 s bins with errors added in quadrature.}
\label{fig:source_lightcurves}
\end{figure*}

\begin{table}
\def\arraystretch{1.2}
\centering
\caption{Results for the KS, Kuiper, and Gregory-Loredo tests for variability. The middle column shows the p-values for the former two tests and the odds ratio for the latter. The third column indicates the interpretation of the tests regarding the X-ray variability of each source.}
\begin{tabular}{lll}
\hline
& LHS~1140 \\
\hline
Kolmogorov-Smirnov
    & $p = 4.0 \times 10^{-3}$ 
    & maybe variable \\
Kuiper
    & $p = 9.2 \times 10^{-4}$
    & variable \\
Gregory-Loredo
    & $O = 4.6$
    & variable \\
\hline
& Background \\
\hline
Kolmogorov-Smirnov
    & $p = 0.53$
    & not variable \\
Kuiper
    & $p = 0.49$
    & not variable \\
Gregory-Loredo
    & $O = 6.0 \times 10^{-3}$
    & not variable \\
\hline
& Contaminant \\
\hline
Kolmogorov-Smirnov
    & $p = 5.5 \times 10^{-2}$
    & not variable \\
Kuiper
    & $p = 7.3 \times 10^{-2}$
    & not variable \\
Gregory-Loredo
    & $O = 5.0 \times 10^{-2}$
    & not variable \\
\hline
\label{table2}
\end{tabular}
\end{table}

\begin{figure*}[t]
\centering
\includegraphics[width=\textwidth]{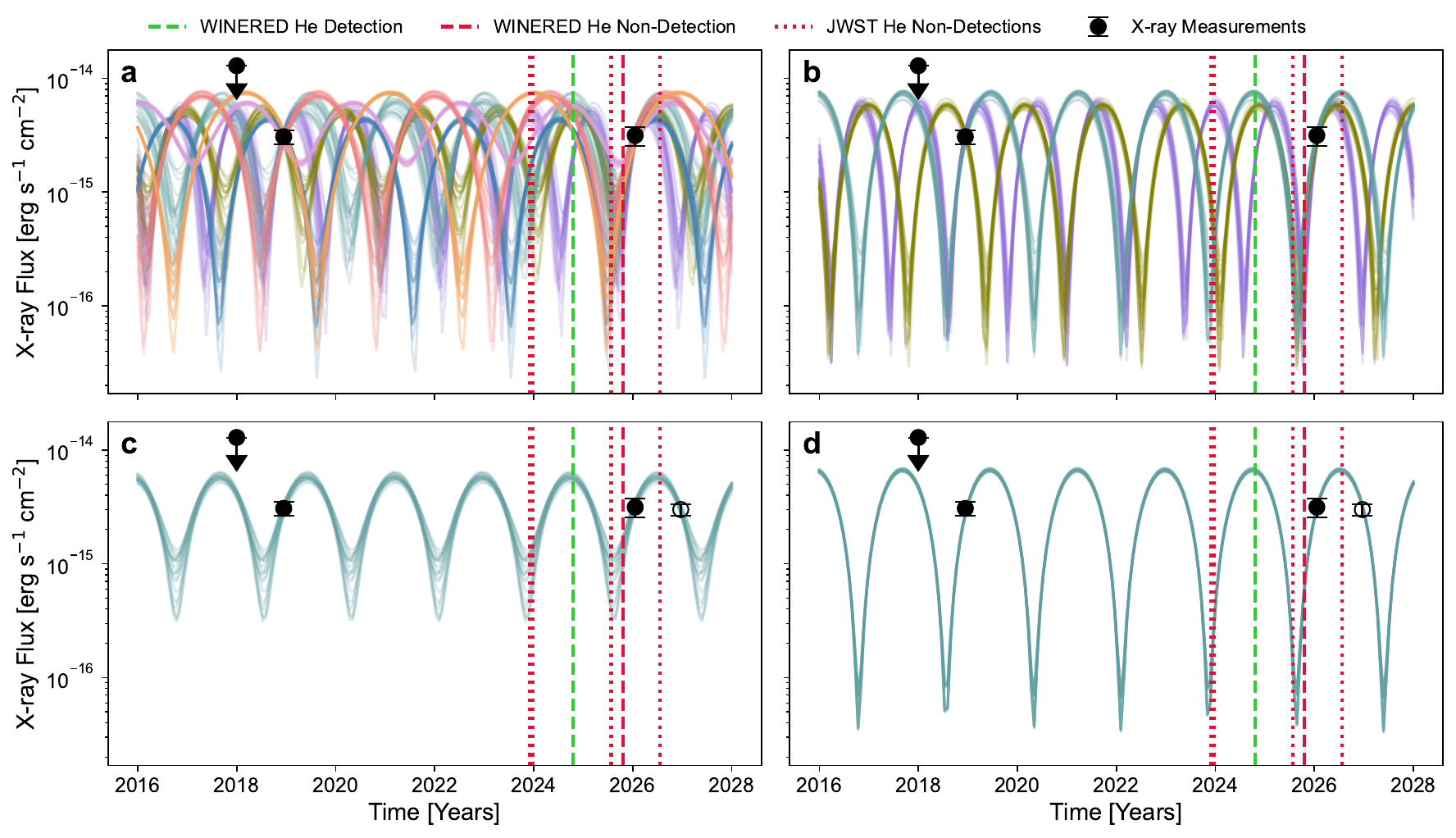}
\caption{X-ray activity cycles consistent with the X-ray measurements of LHS~1140 and the He observations of LHS~1140b. Measured X-ray fluxes are plotted in black. The vertical green and red lines indicate the detection and non-detections of metastable He from LHS~1140b, respectively. Sine curves are plotted from a grid search of periods between 1 and 20 years. The curves were filtered such that the X-ray flux at the time of the WINERED He non-detection on Sep. 28, 2025 (red dashed line) must be 3$\times$ lower (panels a and c) or 10$\times$ lower (panels b and d) than that at the time of the He detection made on Sep. 23, 2025 (green dashed line). This constraint was not enforced for the JWST non-detections (red dotted lines). The sine solutions were also filtered to pass through the measured X-ray points. A hypothetical X-ray measurement on Dec. 2026, during the next visible epoch with \textit{XMM-Newton}, is shown as an open circle in panels c and d. Panels c and d show cycles consistent with this measurement if it reveals an X-ray flux similar to the previous measurements.}
\label{fig:cycles}
\end{figure*}

The best-fit spectral models are shown in Figure \ref{fig:spectra}. We estimate an X-ray flux for LHS~1140 of $3.15^{+0.53}_{-0.65} \times 10^{-15}$\,erg\,s$^{-1}$\,cm$^{-2}$ in the 0.25--2.0\,keV energy range and $3.18^{+0.54}_{-0.66} \times 10^{-15}$\,erg\,s$^{-1}$\,cm$^{-2}$ in the 0.25--10.0 keV energy range. The posterior distributions for the five parameters in the two-APEC model are shown in Figure \ref{fig:corner} and the results are listed in Table \ref{table1}. Our constraints on the normalizations and temperatures for both the soft and hard components of the model are improved compared to the X-ray constraint from the 2018 measurement presented by \cite{Cherubim_2026}. This is due to a higher overall photon count and better separation from the neighboring contaminant. We detect 79 counts in 50 ks in 2026 vs. 25 counts for 31 ks in 2018 in the PN source apertures. When we shrink the 10 arcsec aperture from 2026 to 6.5 arcsec to match that in 2018, we detect 41 counts, consistent with the 2018 count rate.

The raw, non-background subtracted light curves for the EPIC PN and OM exposures are shown in Figure \ref{fig:source_lightcurves}. The OM light curves do not show significant variability in the V filter (centered on 542\,nm) nor in the UVW1 filter (centered on 291\,nm). This is supported by Kolmogorov–Smirnov (KS) tests performed in the \textit{XMM-Newton} Pipeline Processing System (PPS) data products, which suggest that the visible and UV fluxes are constant. 

The X-ray flux on the other hand appears to vary, with an apparent peak roughly centered in the light curve. Following the convention of the Chandra source catalog \citep{Evans_2024}, we performed three statistical tests to quantify the probability that X-ray observations indicate variability. 
These tests include the KS test, Kuiper test, and Gregory-Loredo test \citep{kolmogorov_1933, smirnov_1936, kuiper_1960, Gregory_1992}. We performed these tests for variability on the light curves for the source, background, and neighboring contaminant. We performed the tests on the PN data alone (79 counts) and the combined data from PN, MOS\,1, and MOS\,2 detectors (110 counts). The tests on separate datasets agreed and the evidence for variability increased when the data were pooled. The results for the tests on the pooled data are shown in Table\,\ref{table2}.

Each test quantifies the confidence with which the null hypothesis of constancy can be rejected. For the KS and Kuiper tests, p-values $\leq 0.0026$ indicate that constant flux can be rejected with $\geq 3 \sigma$ confidence, which we choose as our significance threshold. For the Gregory-Loredo test, $O$ indicates the odds ratio of obtaining the observed distribution vs. obtaining a constant distribution. Hence, $O > 1$ favors variability. We adopt the Chandra source catalog convention that $O \geq 2$ is strong evidence for variability. We find no evidence for variability in the background or contaminant apertures (all light curves shown in Figure \ref{fig:all_lightcurves}). In the Kuiper and Gregory-Loredo tests, we find strong evidence for variability of the source X-ray flux.
In the KS test, we find slightly weaker evidence for variability, corresponding to 2.9\,$\sigma$ significance. We note that while the KS and Kuiper tests are very similar, the KS test is less sensitive to detecting the two-sided type of variability, with an ``S'' shaped CDF, which we observe (Figure \ref{fig:kuiper}).

To quantify the amplitude of source X-ray variability, we binned the light curve so each bin includes at least one count (as in Figure \ref{fig:source_lightcurves}), and divided the maximum count rate ($F_\mathrm{max}$) by the minimum count rate ($F_\mathrm{min}$). To determine uncertainties, we drew 100,000 random samples from Gaussian distributions centered on $F_\mathrm{max}$ and $F_\mathrm{min}$, with standard deviations corresponding to their error bars. The median and 16/84\% confidence intervals are $F_\mathrm{max}$/$F_\mathrm{min} = 8.5^{+9.4}_{-4.4}$.


The source light curve appears to peak toward the center of the $\sim$50 ks observing window.
To investigate the nature of this feature, we fit four models to the data combined across the three EPIC detectors: a flat line, a two-level box model, a sinusoid, and a standard fast rise exponential decay (FRED) flare model. Using a chi-squared fit, the results are sensitive to the binning, though a flat line is always rejected ($\chi^2\ \mathrm{p} \approx 0.004$), and the sinusoid model tends to be favored. Given the sensitivity to binning, we fit the unbinned data using C-statistics \citep{cash1979} as in our spectral analysis. We find that the box model is preferred with differences in Bayesian information criteria ($\Delta$BIC) of 7.21, 6.31, and 10.66 for the flat, sinusoid, and FRED models, respectively (Figure \ref{fig:epic_models}). Following the goodness-of-fit approach of \texttt{XSPEC}, we find that the flat line is rejected (p = 0.005) and the box, sinusoid, and FRED models are all acceptable (p = 0.54, 0.17, and 0.22). We therefore cannot distinguish between a flare or sustained enhancement, which is unsurprising in a low count regime.

To further test the stellar flare hypothesis, we computed hardness ratios, which compare fluxes of low- and high-energy X-rays (Figures \ref{fig:HR1} and \ref{fig:HR2}). Flares are associated with high-energy activity in the stellar corona, and are typically associated with elevated hardness ratios relative to inactive periods (\cite{Pye_2025}, see Figures 12 -- 15). The GTIs from the overall exposure were split into thirds, each equal in duration. The event lists for the two outer segments were concatenated into a single event list. Two light curves were then generated as described in Section\,\ref{sec:analysis}, representing the middle segment and concatenated outer segments. We find no significant evidence that the hardness ratios of these segments differ from each other, or from that of the overall light curve. While we find no statistically significant evidence for a time-variable hardness ratio, the measurement precision is too low to confidently rule out a flare. Larger flares may exhibit corresponding flares in the UVW1 bandpass, which we do not observe \citep{Pye_2025}. To our knowledge, no flares have been observed at any wavelength for this star, which is old and quiescent \citep{Dittmann_2017, spinelli2019, Medina_2022}.

We repeated the variability tests for the \textit{XMM-Newton} data collected in 2018 \citep{Cherubim_2026} for the source, contaminant, and background apertures. The results are shown in Table \ref{table3} and the light curves are shown in Figure \ref{fig:all_lightcurves_2018}. We find no significant evidence for variability in any test for the source. Two tests suggest the background is variable, and all tests suggest the contaminant is variable. These data are less conclusive than the 2026 data given that the 2018 time series is interrupted by frequent background particle flaring events and is therefore more sparse. Bootstrapping our 2026 detector-combined counts (110) down to 34 counts (the 2018 total), we estimate a $\sim 50 \%$ probability of detecting the 2026 variability via the Kuiper test, ignoring the fact that the background and contaminant may contribute variable noise.

\section{Discussion}

\subsection{Variable XUV flux as an explanation for variable He absorption}

\cite{Cherubim_2026} showed that variations in XUV flux of a factor of $\sim 10 \times$ and/or exosphere temperature changes of $\sim 1000$ K could explain the He absorption detection and non-detection observed in 2024 and 2025, respectively. We show that the X-ray flux of LHS~1140 varied by a factor of $8.5^{+9.4}_{-4.4} \times$ within the 18-hour observing window. Assuming a sound speed of $c_s = \sqrt{kT/\mu} = 3270$ m s$^{-1}$, where $k$ is the Boltzmann constant, $T = 5160$ K is the estimated outflow temperature from \cite{Cherubim_2026}, and $\mu = 4$ amu for a He atmosphere, we find that a sound wave takes $\lesssim 1$ hour to propagate a distance of 1 planetary radius through LHS~1140b's atmosphere. This suggests that the readjustment timescale for the mass-loss rate to respond to a changing XUV flux is less than the 18-hour observing time. This is consistent with 3D hydrodynamic simulations that show escaping atmospheres respond to XUV flares within $\sim$ 2 hours \citep{Wang_2021a} and with observations of Earth's thermosphere response to solar flares \citep{Sutton_2006, Qian_2011}. We conclude that short-timescale stellar XUV variability is a plausible explanation for the metastable He variability observed during the transit of LHS~1140b.

\subsection{Physical explanations for the X-ray variability}
\label{sec:xray_mechanisms}

Stellar X-ray emission is generated from plasma in the corona that is heated due to magnetic activity. Fully convective M stars like LHS~1140 have complex magnetic activity that is poorly understood \citep{Lehmann_2024, Henry_2024}. Stellar age is negatively correlated with X-ray flux and X-ray cycle amplitudes, presumably because older stars are not saturated with X-ray emitting coronal magnetic structures, allowing for more X-ray flux variability \citep{Coffaro_2022}. LHS~1140 is an old, quiescent star \citep{Dittmann_2017, Medina_2022}, and so it may exhibit short-timescale X-ray variability due to stochastic magnetic activity.

The smooth change in flux that we observe does not resemble flares observed with \textit{XMM-Newton}, which instead spike from quiescence then gradually decay in the X-ray \citep{Pye_2025, Pillitteri_2026}, but we do not rule out a small flare.
If the observed X-ray variability did result from a flare, it is difficult to infer how much the total XUV flux changes based on the X-ray measurement alone. For solar flares, the EUV and X-ray flux peaks nearly simultaneously \citep{Thiemann_2017}. EUV emission dominates over X-ray emission in the energy budget of solar flares, but the increase in EUV flux relative to quiescence is less than that observed for the X-ray flux since quiescent EUV flux is higher. The picture is complicated by the difference in X-ray vs. EUV emission timescales, as EUV emission persists far longer after peak X-ray emission, oftentimes enhanced by the EUV late phase \citep{Greatorex_2026}. We leave a quantitative analysis of the total XUV variability that may drive escape for future studies.

\subsection{Ruling out long-term X-ray activity cycles}
\label{sec:xray_activity}

Our measured, integrated X-ray flux is fully consistent with the 2018 measurement. This suggests that long-term X-ray flux variability may be low, though it is possible that the two X-ray measurements occurred in the same phase of a sinusoid-like coronal activity cycle, which has been previously observed for other stars \citep{Hathaway_2015, Wargelin_2024, Pillitteri_2026}. In Figure \ref{fig:cycles}, we show results of a grid search for sinusoid models consistent with the X-ray measurements and with periods between 1 and 20 days and fractional amplitudes between 1 and 99\%. We define cycle amplitudes as $F_\mathrm{ max}/F_\mathrm{min}$, as in \cite{Pillitteri_2026}. For each sinusoid solution, we assume the X-ray flux in 2025 was three times lower (Figure \ref{fig:cycles}a and c) or ten times lower (Figure \ref{fig:cycles}b and d) than in 2018, motivated by the He line models consistent with the He non-detection \citep{Cherubim_2026}. These criteria assume that the XUV flux was lower during the non-detection in 2025 compared to 2024 when He was detected.

The red dotted lines in Figure \ref{fig:cycles} indicate four \textit{JWST} observations that include coverage of the metastable He line with NIRISS. He absorption corresponding to $\sim$60--80\% of that observed in 2024 (i.e., 0.75--1.0\% excess absorption convolved to WINERED resolution) has been ruled out for those observations \citep{Radica_2026, Bennett_2026, Gressier_2026}. These constraints are not used to filter sine solutions in our grid search, though three of four \textit{JWST} observations are consistent with most solutions. If all \textit{JWST} constraints are included, no solutions remain. We also show an upper limit on the X-ray flux from \cite{spinelli2023}, which does not help constrain potential activity cycles.


The next window of visibility of LHS~1140 with \textit{XMM-Newton} is in December, 2026. If an additional measurement in this window is consistent with the previous measurements, this would drastically reduce the viable sine solutions to those with periods less than 2 years (Figure \ref{fig:cycles}c and d). In this case, ignoring the He escape measurements, the statistically favored solution would be a linear fit to the X-ray measurements and an X-ray activity cycle would not explain the He escape variability. 

Two M stars have observed X-ray activity cycles, L\,98-59 (an old, inactive M3 star), and Proxima Centauri (an old, highly active M5.5 star), with cycle amplitudes of $\sim10 \times$ and $\sim 1.5 \times$ and periods of 2 years and 8 years, respectively \citep{Wargelin_2024, Pillitteri_2026}. Six other X-ray cycles have been observed for Sun-like stars, all with amplitudes less than $10 \times$, and periods typically greater than two years \citep{Robrade_2012_61CygA, Sanz-Forcada_2013_iHor, Robrade_2016_alphacenAB, Orlando_2017_HD81809, Coffaro_2020}. Older stars like LHS~1140 tend to have higher X-ray cycle amplitudes, possibly due to the allowed variability in coronal spots on a relatively inactive surface \citep{Coffaro_2022}. All but three models in Figure \ref{fig:cycles}a have cycle amplitudes greater than an order of magnitude, which would make them outliers, and those three have short periods of 1.6, 1.8, and 2.2 years.

Long-term monitoring of chromospheric Ca\,{\sc ii} and optical continuum emission shows that M dwarfs typically exhibit activity cycles ranging from 2 to 20+ years \citep{Suarez_2016, Alonso_2019, Mignon_2023}. While X-ray activity cycles vary with chromospheric activity cycles measured via the S-index for Sun-like stars \citep{Coffaro_2022}, X-ray cycles for M stars are less well understood, as only two have been observed, for L\,98-59 and Proxima Centauri \citep{Pillitteri_2026, Wargelin_2024}. There is no correlation between stellar rotation period or age and chromospheric activity cycle period for M dwarfs, likely due to irregular magnetic surface activity \citep{Savanov_2012, Suarez_2016, Alonso_2019}. It is therefore difficult to predict LHS\,1140's X-ray variability and whether an X-ray activity cycle may correlate with chromospheric activity, further motivating additional X-ray/UV measurements for LHS~1140.

Note that our models only explore the sinusoidal coronal activity cycle hypothesis and ignore stochastic variability. If stochastic variability dominates, simultaneous stellar XUV and in-transit metastable He observations are truly necessary to establish a link between variable XUV flux and variable He escape.

\subsection{Alternative explanations for the He absorption variability of LHS~1140b}

It is possible that atmospheric escape was in fact occurring in 2025 at a similar rate as that measured in 2024, but metastable He was insufficiently populated to be detected in the outflow. If the sinks for metastable He, primarily ionization, overwhelm the sources, the outflow could evade detection. 
The primary source of metastable He is neutral He ionization by extreme-ultraviolet photons of wavelengths $\lambda < 504$ \AA\ followed by recombination.
Our data do not cover 504 \AA\ and the observed X-ray flux is the best available proxy. The secondary source of metastable He is collisional excitation, which depends on X-ray heating. The observed X-ray variability suggests that population of metastable He may vary in the atmosphere of LHS~1140b. The primary sink for metastable He is ionization by photons of wavelength $\lambda < 2583$ \AA, which is covered by our OM observations with the UVW1 filter. The lack of UV variability in our observations suggests that the stellar energy flux available to destroy atmospheric metastable He is roughly constant on hours-long timescales.
 
Besides XUV flux variability, other less well understood physical processes might explain the observed He variability. 
Beyond LHS~1140b, \cite{Cherubim_2026} discuss several other cases of variable atmospheric escape of neutral H and metastable He from other exoplanets.
Atmospheric escape is a complex process involving fluid dynamics in three spatial dimensions, which makes the interpretation of metastable He transmission spectra difficult. For example, stellar wind variability and shear instabilities could affect the atmospheric mass loss rate and manifestation of the He line in transmission spectra \citep{Wang_2021}. Additionally, \cite{Carolan_2020} and \cite{Schreyer_2024} showed that strong stellar winds and planetary magnetic fields can dramatically reduce atmospheric outflow rates and/or diminish the metastable He absorption depth.

\section{Conclusion}

We present first results from an XMM-Newton program that will monitor the XUV flux of LHS~1140 over a $\sim$1.5 year baseline. The goal is to determine whether stellar XUV variability is correlated with the observed variability in the metastable He line depth consistent with a planetary atmosphere reported by \cite{Cherubim_2026}. In this work, we find robust evidence for X-ray variability with a dynamic amplitude of $8.5^{+9.4}_{-4.4} \times$ over an $\sim$18-hour period. This magnitude of variability is roughly consistent with that predicted to explain the He detection and non-detection in 2024 and 2025, respectively, reported by \cite{Cherubim_2026}. We do not detect any variability in the visible or UV flux, nor in the hardness ratio of the X-ray emission. This suggests that large flares are unlikely and that the sources for metastable He in the atmosphere of LHS~1140b are likely to vary while the primary sink may not. 

We measure an overall mean X-ray flux consistent with that measured in 2018, suggesting the X-ray flux may be stable on years-long timescales. While it is possible that we coincidentally observed similar X-ray fluxes in the same phase of a coronal activity cycle that may explain the planetary He absorption variability, we conclude that this is unlikely based on previously observed X-ray cycles for similar stellar types. Alternatively, variability on hours-long timescales remains a plausible explanation for the He variability, whether from changes in mass-loss rate or metastable He population. Simultaneous XUV/near-infrared transit observations are needed to break the current degeneracies and help elucidate whether XUV flux variability or other mechanisms (e.g., shear instabilities, stellar wind variability, magnetic field interactions, metastable He depopulation) dominate the observed He absorption variability.

LHS~1140b is a nearby, rocky exoplanet orbiting in the habitable zone of a quiet star. As such, it remains an ideal astrobiological laboratory and provides a unique opportunity to study the connection between XUV irradiation and atmospheric escape. In light of the X-ray measurement presented here, the X-ray irradiation of LHS~1140b appears to be stable on long timescales and relatively low -- just 3-13 times that of Earth depending on the Solar activity cycle phase: the lowest of all well-characterized super Earths transiting nearby M dwarfs. Taken together with previous results that show no flares in the X-ray or UV, relatively low far- and near-UV fluxes \citep{Spinelli_2019}, the presence of an atmosphere \citep{Cherubim_2026}, and an equilibrium temperature of 230 K, LHS~1140b is a promising world for studying habitability outside the solar system.

\begin{acknowledgments}

We thank Anna Ruth Taylor, Ruth Murray-Clay, and Shreyas Vissapragada for helpful discussions.
C. C. and T.C. were supported by NASA through the NASA Hubble Fellowship grants HST-HF2-51601.001-A and HST-HF2-51527.001-A, respectively, awarded by the Space Telescope Science Institute, which is operated by the Association of Universities for Research in Astronomy, Inc., for NASA, under contract NAS5-26555.
We recognize support from NASA grant 80NSSC26K1187.

\end{acknowledgments}




%
\facilities{XMM-Newton}

\software{  
          XSPEC \citep{arnaud1996XSPEC},
          BXA \citep{buchner2016-bxa},
          SAS \citep{gabriel2004-sas},
          Astropy \citep{2013A&A...558A..33A,2018AJ....156..123A,2022ApJ...935..167A},
          SciPy \citep{scipy},
          NumPy \citep{numpy},
          Matplotlib \citep{matplotlib}
          }


\appendix

\section{Supplementary Figures and Table}

\setcounter{figure}{0}
\renewcommand{\thefigure}{S\arabic{figure}}

\setcounter{table}{0}
\renewcommand{\thetable}{S\arabic{table}}

\FloatBarrier

\begin{table}[!htb]
\def\arraystretch{1.2}
\centering
\caption{Results for the Kolmogorov-Smirnov, Kuiper, and Gregory-Loredo tests for variability for the 2018 dataset. The middle column shows the p-values for the former two tests and the odds ratio for the latter. The third column indicates the interpretation of the tests regarding the X-ray variability of each source.}
\begin{tabular}{lll}
\hline
& LHS~1140 \\
\hline
Kolmogorov-Smirnov
    & $p = 0.69$
    & not variable \\
Kuiper
    & $p = 0.39$
    & not variable \\
Gregory-Loredo
    & $O = 0.32$
    & not variable \\
\hline
& Background \\
\hline
Kolmogorov-Smirnov
    & $p = 4.9 \times 10^{-2}$
    & not variable \\
Kuiper
    & $p = 2.8 \times 10^{-4}$
    & variable \\
Gregory-Loredo
    & $O = 2.2 \times 10^{2}$
    & variable \\
\hline
& Contaminant \\
\hline
Kolmogorov-Smirnov
    & $p = 5.1 \times 10^{-6}$
    & variable \\
Kuiper
    & $p = 4.0 \times 10^{-5}$
    & variable \\
Gregory-Loredo
    & $O = 1.3 \times 10^{3}$
    & variable \\
\hline
\label{table3}
\end{tabular}
\end{table}

\begin{figure*}
\centering
\includegraphics[width=1.0\textwidth]{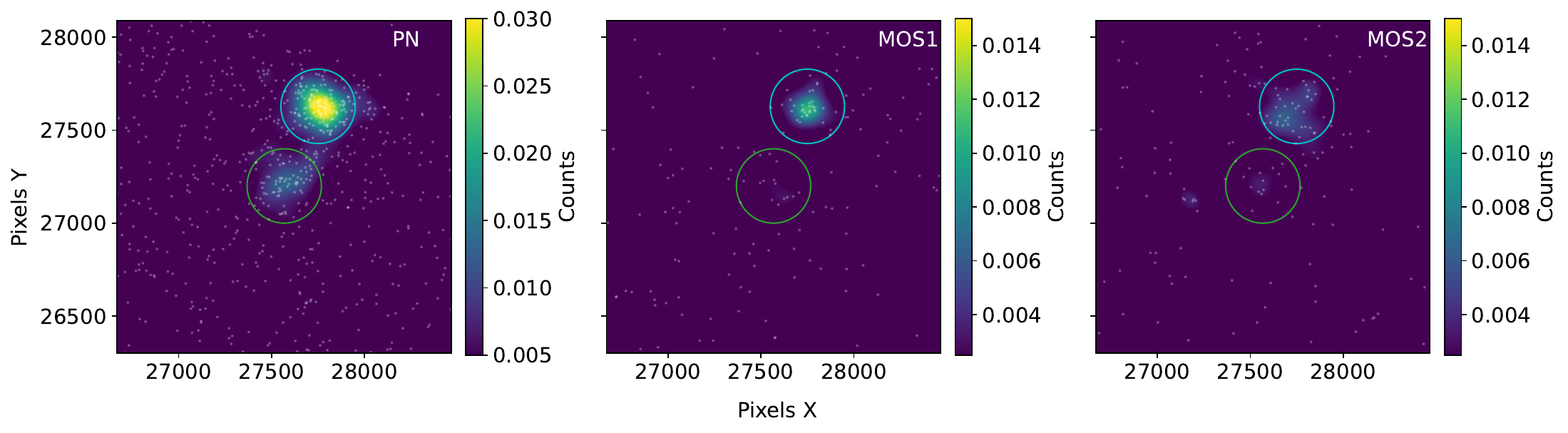}
\caption{Apertures for the source (green circle) and contaminant (blue circle) on the PN, MOS1, and MOS2 detectors. Gray points show detected events. Colors show event counts convolved with the instrument point spread function. Axes show physical detector units.}
\label{fig:apertures}
\end{figure*}

\begin{figure*}
\centering
\includegraphics[width=1.0\textwidth]{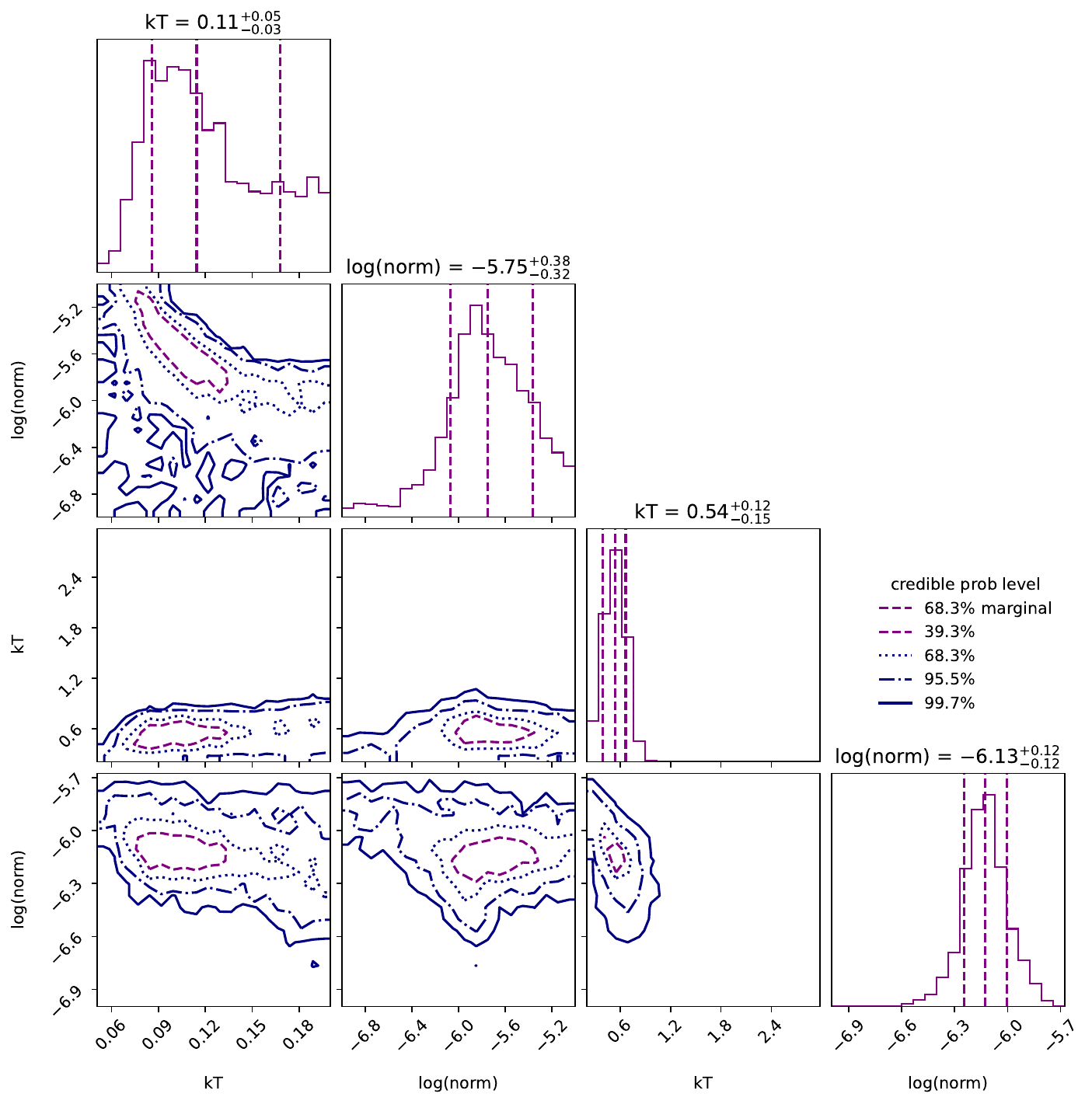}
\caption{Posterior distributions from the BXA fit of the four fitted parameters in the APEC optically-thin plasma model: two temperatures ($kT$, Table \ref{table1}) and two normalizations (norm).}
\label{fig:corner}
\end{figure*}

\begin{figure*}
\centering
\includegraphics[width=1.0\textwidth]{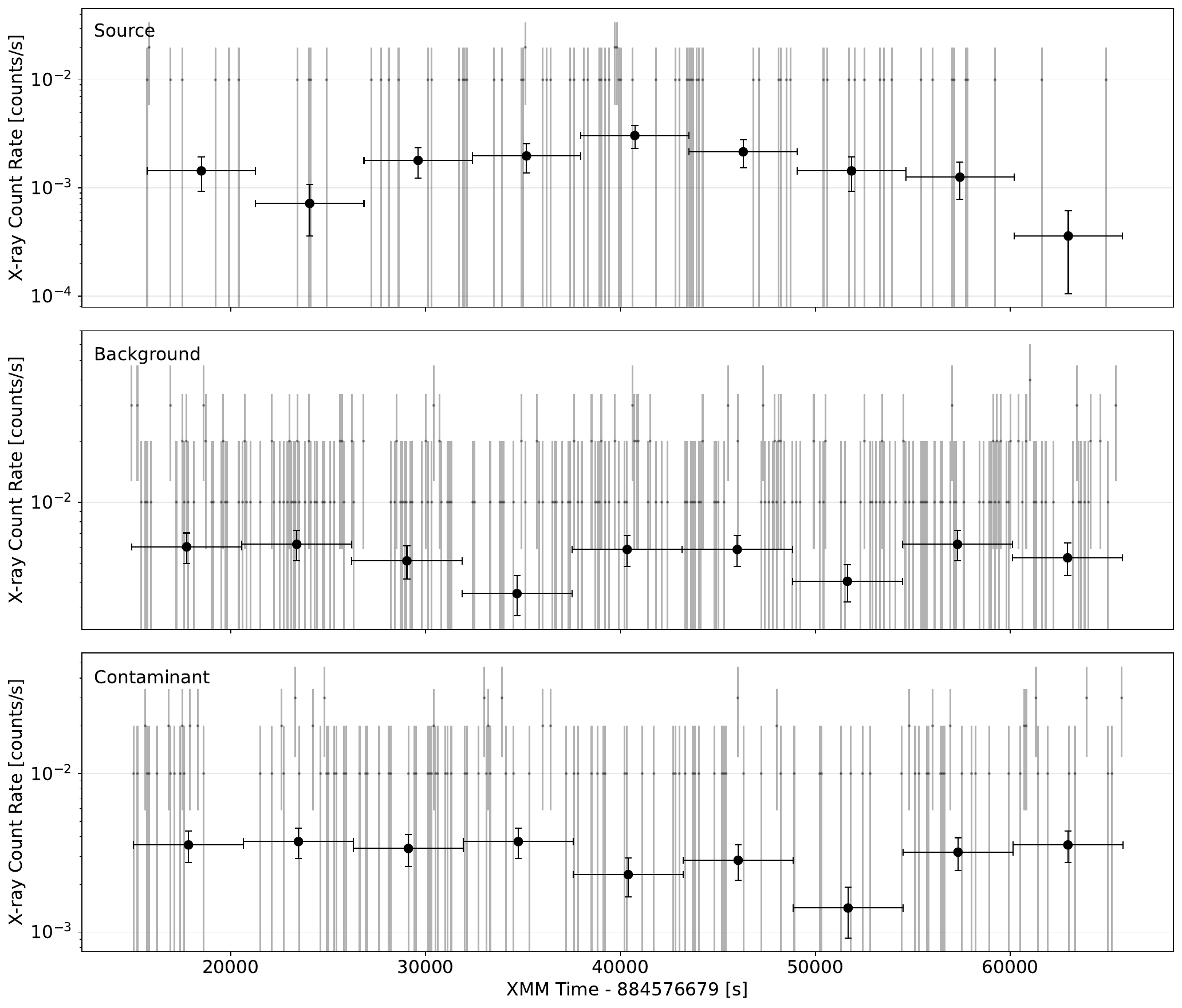}
\caption{Light curves for source (top) background (middle) and contaminant (bottom) showing photon counts over time for the 2026 data. Gray points show data binned to 100 s with error bars generated from the XMM-Newton Science Analysis System pipeline. The absence of lines indicates zero counts. Black points show 5600 s bins, which were chosen to include a minimum of one count for the source aperture. Error bars show errors added in quadrature.}
\label{fig:all_lightcurves}
\end{figure*}

\begin{figure*}[t]
\centering
\includegraphics[width=0.5\textwidth]{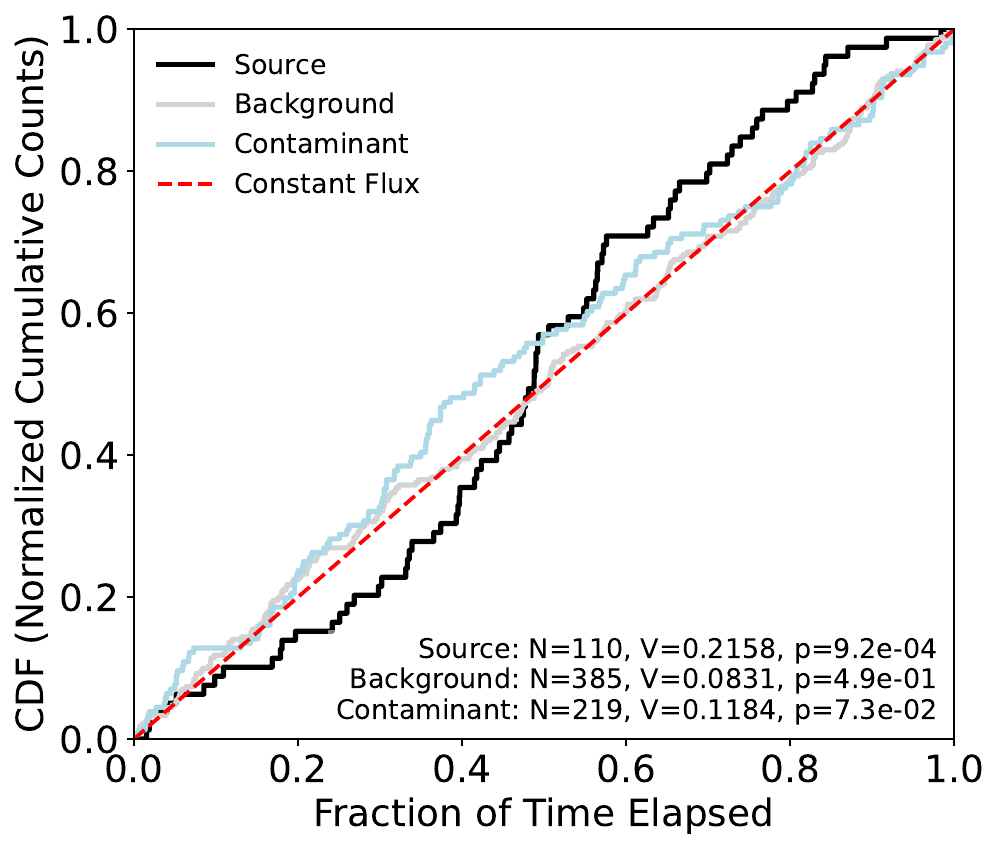}
\caption{Kuiper test for X-ray variability. Black, gray, and blue points show the cumulative distribution function (CDF) computed from the source, background, and contaminant apertures, respectively. CDFs were constructed by summing counts in time. Each CDF was compared to a constant flux (red dashed line). The resulting Kuiper scores and p-values are shown. P-values indicate the probability that each observed CDF is consistent with a constant flux. The results are summarized in Table \ref{table2}.}
\label{fig:kuiper}
\end{figure*}

\begin{figure*}[t]
\centering
\includegraphics[width=0.7\textwidth]{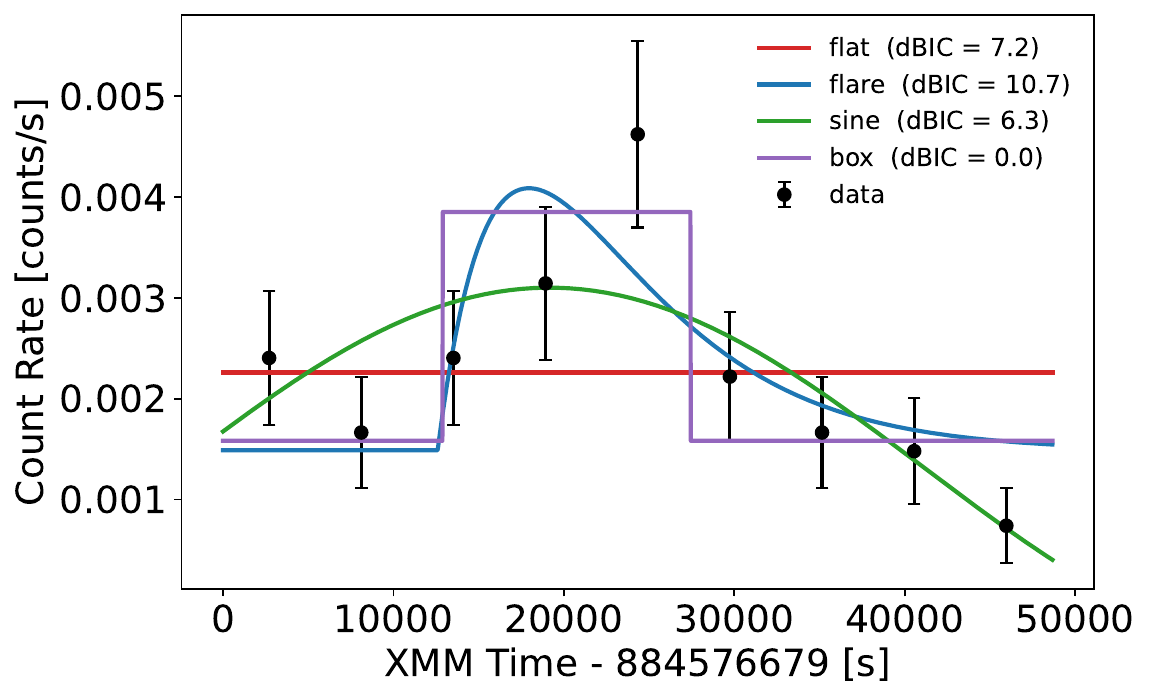}
\caption{Model fits to unbinned, combined EPIC event lists (110 photons). The flare model is a fast rise exponential decay model with 5 parameters. The sine model is a sine wave with 4 parameters. The box model is a two-level profile with 4 parameters. We calculated the goodness of fit of each model to the unbinned data using the C-statistic \citep{cash1979}. We then calculated the BIC values as $-2\mathrm{ln}\hat{L}\ +\ k\mathrm{ln}(n)$, where $\hat{L}$ is the C-statistic maximum likelihood value, $k$ is the number of model parameters, and $n$ is the sample size (110 photons). The reported differences in BIC, $\Delta$BIC, are relative to the best fitting model: the box model. A higher dBIC indicates a worse fit. Data are shown in nine bins only for visualization; no models were fit to binned data.}
\label{fig:epic_models}
\end{figure*}

\begin{figure*}[t]
\centering
\includegraphics[width=0.5\textwidth]{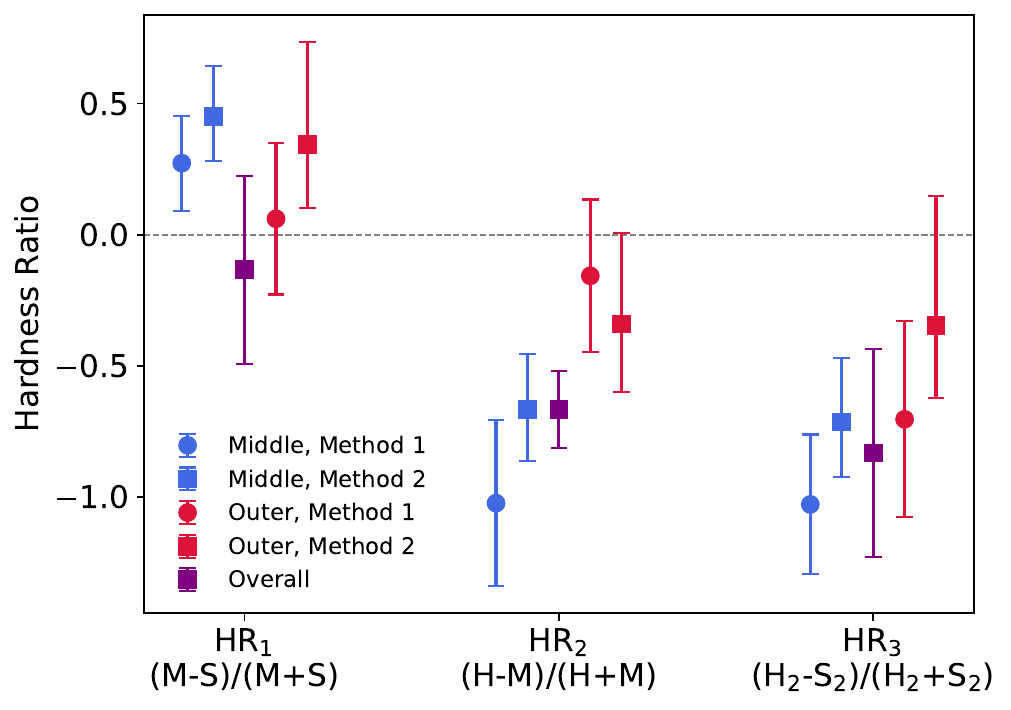}
\caption{Hardness ratios for the middle (blue), outer (red), and overall (purple) segments of the PN X-ray light curve. The overall light curve was split into thirds, each equal in duration. The outer two segments were concatenated and represent the ``outer'' segment. The middle third represents the ``middle'' segment. The ``overall'' segment includes all continuous counts. For Method 1, hardness ratios were computed from raw photon counts. For Method 2, they were computed from the best fitting fluxes from the BXA MCMC analysis. The energy bands are as follows, H: 1.0 - 2.0 keV; M: 0.5 - 1.0 keV; S: 0.2 - 0.5 keV; H$_\mathrm{2}$: 1.0 - 10.0 keV; S$_\mathrm{2}$: 0.2 - 1.0 keV.}
\label{fig:HR1}
\end{figure*}

\begin{figure*}[t]
\centering
\includegraphics[width=0.5\textwidth]{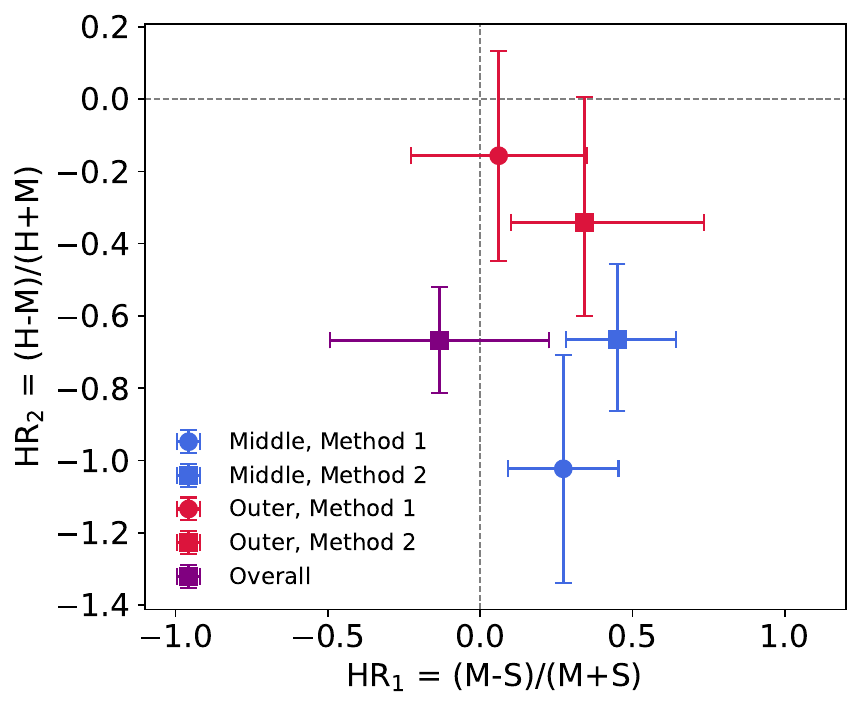}
\caption{Same as Figure \ref{fig:HR1}, but only HR$_\mathrm{1}$ and HR$_\mathrm{2}$. Compare to Figure 4 in \cite{Caramazza_2023}.}
\label{fig:HR2}
\end{figure*}

\begin{figure*}
\centering
\includegraphics[width=1.0\textwidth]{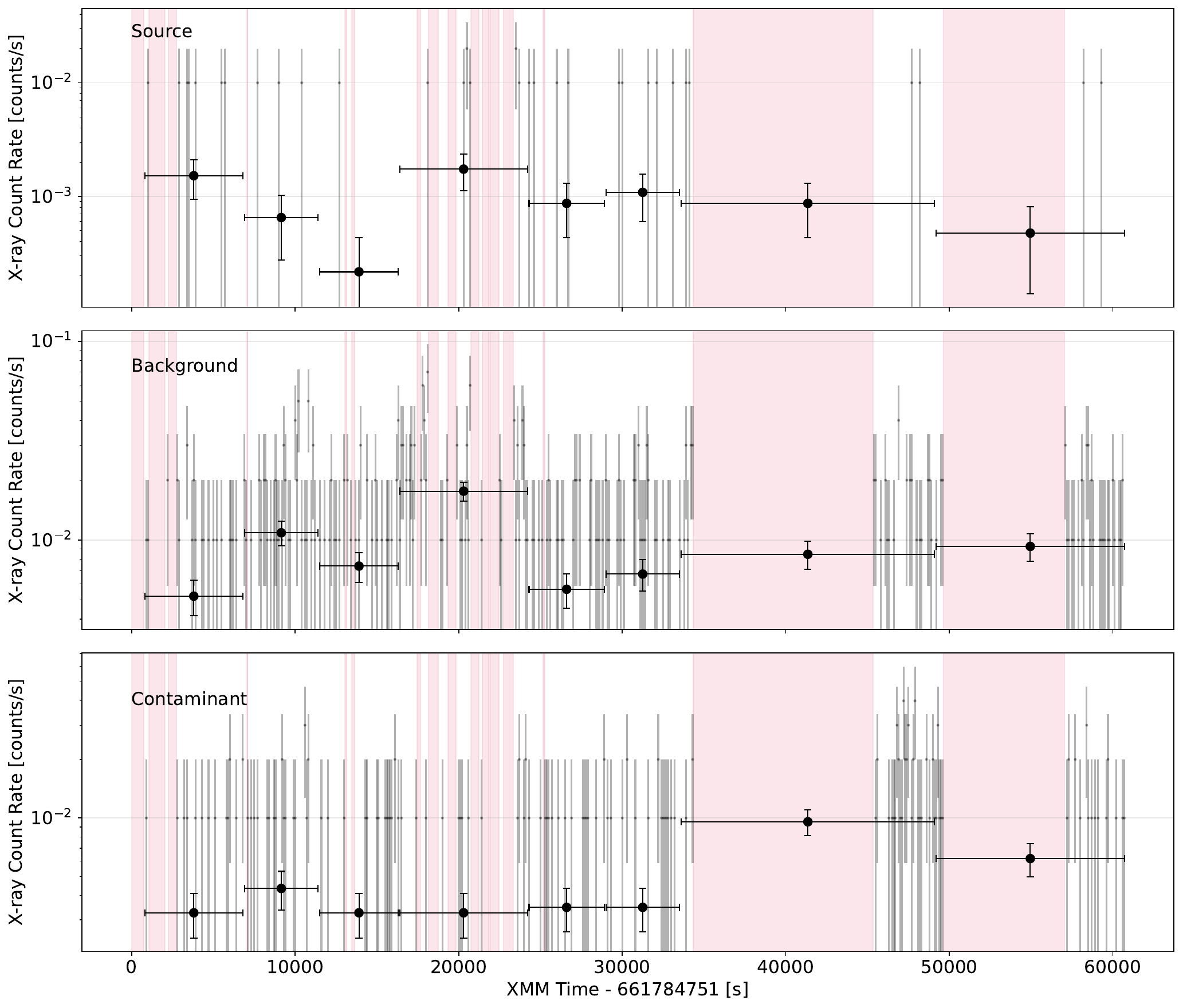}
\caption{Same as Figure \ref{fig:all_lightcurves} but for the 2018 data. Black points show 4600 s bins, which include a minimum of one count. Shaded red regions show non-GTI intervals contaminated by high background events, as in Figure \ref{fig:source_lightcurves}.}
\label{fig:all_lightcurves_2018}
\end{figure*}

\clearpage
\bibliography{references}{}
\bibliographystyle{aasjournalv7}



\end{document}